\documentclass[showpacs,twocolumn,preprintnumbers,amsmath,amssymb,epsf]{revtex4}
 \newcommand{\eins}{\mbox{$1 \hspace{-1.0mm}  {\bf l}$}}
 \newcommand{\ket}[1]{ | \, #1  \rangle}
 
 \newcommand{\bra}[1]{ \langle #1 \,  |}
 
 \newcommand{\proj}[1]{\ket{#1}\bra{#1}}
\def\map#1{{\cal{#1}}}
\def\set#1{{\cal #1}}
\def\sH{\set{H}}
\def\Cmap{\map{C}}
\def\Imap{\map{I}}
\def\){\rangle\!\rangle}\def\({\langle\!\langle}
\def\Tr{\operatorname{Tr}}\def\d{\operatorname{d}}
\begin{document}
\title{Optimal phase covariant cloning for qubits and qutrits}
\author{Giacomo Mauro D'Ariano}
\email{dariano@unipv.it}
\altaffiliation[Also at ]{Department of Electrical and Computer
Engineering, Northwestern University, Evanston, IL  60208}
\author{Chiara Macchiavello}
\email{macchiavello@unipv.it}
\affiliation{{\em Quantum Optics and Information Group},
Istituto Nazionale di Fisica della Materia,
Unit\`a di Pavia}
\homepage{http://www.qubit.it}
\affiliation{Dipartimento di Fisica ``A. Volta'', via Bassi 6, I-27100 Pavia, Italy}
\date{\today}
\begin{abstract}
We consider cloning transformations of equatorial qubits
$\ket{\psi_\phi}=\tfrac{1}{\sqrt2}(\ket{0} +e^{i\phi}\ket{1})$ and qutrits
$\ket{\psi_{\phi,\theta}}=\tfrac{1}{\sqrt 3}(\ket{0}+e^{i\phi}\ket{1}
+e^{i\theta}\ket{2})$, with the transformation covariant for rotation 
of the phases $\phi$ and $\theta$. The optimal cloning maps are derived without
simplifying assumptions from first principles, 
for any number of input and output
qubits, and for a single input qutrit and any number of output qutrits.
We also compare the cloning maps for global and single particle fidelities,
and we show that the two criteria lead to different optimal maps.
\end{abstract}
\maketitle
\date{October 12, 2002}
\maketitle

\section{Introduction}

The impossibility of perfectly cloning unknown quantum states selected
from a nonorthogonal set is a typical quantum feature
\cite{no-cloning}, and is the basis of the security of quantum cryptography
\cite{BB84,E91,Ekert}.  In
fact, the potential eavesdropper Eve cannot clone the quantum
state transmitted by Alice, recover it from multiple copies, and
retransmit it undisturbed to the receiver Bob. Eve,
however, can try to realise an approximate cloning 
\cite{BuzekHillery,gima,bem,Werner}
in an optimal way, maximizing the fidelity of the copies with the
original state, and this is a possible eavesdropping strategy. 
The eavesdropping strategies that are known to be optimal so far are
actually based on cloning attacks \cite{bcdm,3dim,mb1}.
Moreover, quantum cloning allows to study the sharing of quantum 
information among several parties and it may be applied also to study the 
security of multi-party cryptographic schemes \cite{multi-qc}. 

Generally the values of the fidelity achieved by optimal cloning 
transformations depends on the set of allowed input states.
In particular, higher fidelities can be achieved for smaller
sets of input states, since the more information about the input is
given, the better the input states can be cloned. More precisely, for
group covariant cloning 
\cite{dalop}---where the set of input states is the orbit of a given
state under the action of a group of unitary transformations---the
smaller is the group the higher is the fidelity averaged over the input
states. 

In this paper we will develop  a thorough analysis of the cloning map that is
optimal for equatorial qubits and qutrits without any simplifying assumptions, 
including that of group covariance and the requirement that the
output of the cloning map has support on a 
symmetric tensor product Hilbert space. As we will see in Sect. II, 
these assumptions can be derived from the form of the fidelity that
one wants to maximize.
More precisely, we will derive with no assumption the optimal quantum
cloning transformation maximizing the fidelity averaged uniformly over
all states of a Bloch sphere equator
\begin{equation}
|\psi_{\phi}\rangle=\tfrac{1}{\sqrt{2}}\left[\ket{0}+e^{i\phi}\ket{1}\right]\;,
\end{equation}
where $\{\ket{0},\ket{1}\}$ represent a basis for a qubit and
the parameter $\phi\in[0,2\pi)$ is the angle between the Bloch vector 
and the $x$-axis, with the equator in the $xy$ plane with the Bloch
sphere. As we will see, such averaged form for the fidelity
automatically leads to the optimal cloning covariant under the
abelian group  
${\mathbf U(1)}$ of phase rotations---the so-called {\em
phase-covariant} cloning \cite{bcdm}. After the first
analysis of Ref. \cite{bcdm}, where only the upper bounds for the
fidelity were derived by exploiting a connection between optimal phase
covariant cloning and phase estimation, in Refs. \cite{dalop} and
\cite{cinesi} a value for the fidelity that breached the bound given in
Ref. \cite{bcdm} was found for the $1\to 3$ cloning, apparently
obtained under the same assumptions \cite{tri}. Then in Ref. \cite{cinesi} a
cloning transformation from an arbitrary number of input copies $N$ to 
an arbitrary number $M$ of output copies was presented,
however, it was proved to be optimal only for $N=1$. In this
paper we will prove that the cloning maps of Ref. \cite{cinesi} are
generally suboptimal for $N>1$ and we will derive the optimal ones for any 
values of $N$ and $M$. In the
derivation of the optimal cloning maps we will use the general method
designed for group-covariant cloning introduced in Ref. \cite{dalop}, which
exploits the correspondence between CP-maps and positive operators.
We also extended our analysis to the case of phase-covariant cloning 
for {\em qutrits}, namely for quantum states with dimension $d=3$. 
Here the covariance group ${\mathbf
U(1)}\times{\mathbf U(1)}$ is still abelian, and describes the
rotation of two different phases.

\par The paper is organized as follows. In Sect. II we describe the general
theory of optimal phase-covariant cloning, giving the definitions of
all relevant quantities in the qubit case, since for the qutrit case
the treatment is strictly analogous. The starting point is the
maximization of a phase-averaged fidelity, which will lead to
a phase-covariant CP-map with output on the symmetric Hilbert space
of the output copies. Then the theory of group covariant cloning of
Ref. \cite{dalop} is shortly reviewed and specialized to the case of
phase-covariance. In Sect. III we derive the
optimal phase-covariant cloning for qubits for any number of input and
output copies, giving the fidelities for all cases. In Sect. IV the
same derivation is given for qutrits with any number of output
copies, starting from a single input copy. Finally, in Sect. V we
conclude with a discussion of the results, and with some open problems and
future perspectives. 

\section{Optimal phase-covariant cloning}

A cloning map is a special kind of {\em quantum channel}, i. e. a
trace-preserving completely positive (CP) map. In the cloning case,
the CP map $\Cmap$ goes from input states in $\sH$ to output states
in $\sH^{\otimes M}$, with the output state invariant under the
permutations of the $M$ output spaces. More generally, if we have
$N>1$ identical copies available, the map goes from an input state 
$\rho^{\otimes N}$ on the input Hilbert space $\sH_{in}$ given by 
the symmetric subspace ${\cal N}\doteq(\sH^{\otimes N})_+$
of the tensor product $\sH^{\otimes N}$ to the output space 
${\cal M}\doteq\sH^{\otimes M}$ with
$M> N$, and with the output state permutation invariant. 
Actually, as we will see in the following, the optimal map itself will 
have the output state restricted to the symmetric subspace 
${\cal M}\equiv(\sH^{\otimes M})_+$, 
even though, generally, permutation invariance of the state does not 
imply that the state has support in the symmetric subspace.
In the following we will denote a cloning map from $N$ to $M$
``copies'' as $\Cmap_{NM}$.

We want to find a cloning map $\Cmap_{NM}$ which minimizes the
following averaged fidelity
\begin{equation}
\begin{split}
\overline{f}[\Cmap_{NM}]&=\int_0^{2\pi}\frac{\d\phi}{2\pi}\,
f[\Cmap_{NM}](\phi),\\
f[\Cmap_{NM}](\phi)&=\Tr[\ket{\psi_\phi}\bra{\psi_\phi}^{\otimes M}
\Cmap_{NM}(\ket{\psi_\phi}\bra{\psi_\phi}^{\otimes N})],\label{fidglob1}
\end{split}
\end{equation}
for {\em equatorial qubit} input states $\ket{\psi_\phi}$ defined as
\begin{equation}
\ket{\psi_\phi}=\tfrac{1}{\sqrt2}(\ket{0} +e^{i\phi}\ket{1}).
\end{equation}
We will call the cloning map {\em phase-covariant} if it satisfies the
following covariance relation
\begin{equation}
\Cmap_{NM}(U_\phi^{\otimes N}\,\rho_{\cal N}\,U_\phi^{\dagger\otimes N})=
U_\phi^{\otimes M}\,\Cmap_{NM}(\rho_{\cal N})\,
U_\phi^{\dagger\otimes M},
\label{phicov}
\end{equation}
where $U_\phi$ is the unitary phase rotation operator
\begin{equation}
U_\phi=\exp\left[\tfrac{i}{2}\phi(1-\sigma_z)\right],
\label{uphi}
\end{equation}
$\sigma_{x,y,z}$ denotes the usual Pauli matrices, and $\rho_{\cal N}$
is any state in ${\cal N}\equiv(\sH^{\otimes N})_+$. In particular,
according to the fidelity in Eq. (\ref{fidglob1}), 
we will consider only input states of the form
\begin{equation}
\rho_{\cal N}=\ket{\psi_0}\bra{\psi_0}^{\otimes N}.\label{rhoN}
\end{equation}
The unitary transformation in Eq. (\ref{uphi}) gives the phase shift
$U_\phi\ket{\psi_{\phi'}}=\ket{\psi_{\phi'+\phi}}$. For qutrits the
situation will be analogous, with input states of the form 
\begin{equation}
\rho_{\cal N}=\ket{\psi_{0,0}}\bra{\psi_{0,0}}^{\otimes N}, 
\end{equation}
where 
\begin{equation}
\ket{\psi_{\phi,\theta}}=\tfrac{1}{\sqrt 3}(\ket{0}+e^{i\phi}\ket{1}
+e^{i\theta}\ket{2})
\end{equation}
denotes an equatorial qutrit state, and in place
of $U_\phi$ we will consider the two-phase rotation operator
$U_{\phi,\theta}$ which achieves the phase shift
$U_{\phi,\theta}\ket{\psi_{\phi',\theta'}}=\ket{\psi_{\phi'+\phi,\theta'+\theta}}$.

Upon defining the rotated map $\Cmap_{NM}^\phi$ as follows
\begin{equation}
\Cmap_{NM}^\phi(\rho_{\cal N})\doteq 
U_\phi^{\dagger\otimes M}
\Cmap_{NM}(U_\phi^{\otimes N}\,\rho_{\cal N}\,U_\phi^{\dagger\otimes
N})U_\phi^{\otimes M},
\end{equation}
from Eq. (\ref{phicov}) we see that covariance of the map $\Cmap_{NM}$ is
equivalent to the identity $\Cmap_{NM}\equiv\Cmap_{NM}^{\phi}$ for
every $\phi$. Since the fidelity $f[\Cmap_{NM}](\phi)$ is linear
versus the cloning map $\Cmap_{NM}$, the averaged fidelity in
Eq. (\ref{fidglob1}) can be also written in the form
\begin{equation}
\overline{f}[\Cmap_{NM}]=\int_0^{2\pi}\frac{\d\phi}{2\pi}\,
f[\Cmap_{NM}^\phi](0)\equiv f[\overline{\Cmap_{NM}^\phi}](0),\label{fidglob2}
\end{equation}
where clearly $f[\Cmap_{NM}^\phi](0)\equiv f[\Cmap_{NM}](\phi)$, and
the averaged map $\overline{\Cmap_{NM}^\phi}$ is obviously defined as
\begin{equation}
\overline{\Cmap_{NM}^\phi}=\int_0^{2\pi}\frac{\d\phi}{2\pi} \Cmap_{NM}^\phi.
\end{equation}
Since, by definition, the averaged map $\overline{\Cmap_{NM}^\phi}$ is
phase-covariant, Eq. (\ref{fidglob2}) simply means that the cloning
map minimizing the averaged fidelity (\ref{fidglob1}) must be itself
covariant. Therefore, finding the optimal cloning map $\Cmap_{NM}$ 
which minimizes the fidelity (\ref{fidglob1}) is equivalent
to find the optimal phase-covariant map $\Cmap_{NM}$ which minimizes
the following fidelity
\begin{equation}
f_{NM}\doteq f[\Cmap_{NM}](0)=\Tr[\ket{\psi_0}\bra{\psi_0}^{\otimes M}
\Cmap_{NM}(\ket{\psi_0}\bra{\psi_0}^{\otimes N})].
\label{fidglob3}
\end{equation}
Moreover, due to orthogonality with the state
$\ket{\psi_0}\bra{\psi_0}^{\otimes M}$, any component 
of the output state $\Cmap_{NM}(\ket{\psi_0}\bra{\psi_0}^{\otimes N})$
which is not supported on the symmetric subspace $(\sH^{\otimes
M})_+$ will give no contribution to the fidelity
(\ref{fidglob3}). Therefore, there will be always an optimal cloning map 
having output on the symmetric space $(\sH^{\otimes M})_+$, and in the
following we can restrict our attention to such maps only, and take
${\cal M}\equiv(\sH^{\otimes M})_+$.
We will also consider for comparison the average single particle fidelity
\begin{equation}
\begin{split}
&F_{NM}=\tfrac{1}{M}{\mbox{Tr}}[(\proj{\psi_0}\eins^{\otimes M-1}
+\eins \proj{\psi_0}\eins^{\otimes M-2}+...\\
&+\eins^{\otimes M-1} \proj{\psi_0})\proj{\psi_0}^{\otimes N}
\Cmap_{NM}(\ket{\psi_0}\bra{\psi_0}^{\otimes N})].
\label{Fse}
\end{split}
\end{equation}
\par As shown in Ref. \cite{dalop}, it is convenient to study
covariant CP maps in terms of invariant positive operators which are
in one-to-one correspondence with CP maps. In the present context this
means to consider the positive operators $R_{NM}$ defined as
\begin{equation}
R_{NM}=\Cmap_{NM}\otimes\Imap_{\cal N} (|I\)\( I|)
\end{equation}
where $\Imap_{\cal N}$ denotes the identity map over ${\cal N}\equiv(\sH^{\otimes N})_+$,
and $|I\)$ is the maximally entangled vector on ${\cal N}\otimes{\cal N}$
\begin{equation}
|I\)=\sum_{n=0}^N\ket{s_{N,n}}\otimes\ket{s_{N,n}},\label{entI}
\end{equation}
$\{\ket{s_{N,n}}\}$ denoting any orthonormal basis for
$(\sH^{\otimes N})_+$, that we conveniently choose as follows
\begin{equation}
\ket{s_{N,n}}=C(N,n)^{-\tfrac{1}{2}} 
\sum_j P_j^{(N)}\ket{\underbrace{00\ldots0}_{N-n}\underbrace{111\ldots1}_{n}},
\label{sbasis}
\end{equation}
where $\{P_j^{(N)}\}$ denote the permutation operators of the $N$ qubits,
and  $C(N,n)$ is the binomial coefficient $C(N,n)\equiv
N!/n!(N-n)!$. As shown in Ref. \cite{dalop}, one can see that $R_{NM}$
is a positive operator on ${\cal M}\otimes{\cal N}$, which is in
one-to-one correspondence with the CP-map 
$\Cmap_{NM}$, with the trace-preserving condition for the map writing
in terms of the operator $R_{NM}$ as follows
\begin{equation}
\Tr_{{\cal M}}[R_{NM}]=I_{{\cal N}}.\label{trcond}
\end{equation}
The map $\Cmap_{NM}$ can be recovered from the positive operator
$R_{NM}$ as follows
\begin{equation}
\Cmap_{NM}(\rho_{\cal N})=\Tr_{{\cal N}}[(I_{{\cal
M}}\otimes\rho_{\cal N}^{\mathrm t}) R_{NM}],\label{CR}
\end{equation}
where $O^{\mathrm t}$ denotes the transposed operator of $O$ with
respect to the same orthonormal basis (\ref{entI}) chosen for the
maximally entangled vector in Eq. (\ref{sbasis}), namely one defines 
$O^{\mathrm t}\doteq (O^\dag)^*$ 
where the {\em complex conjugated} $O^*$ of the operator $O$ is defined as
the operator having complex conjugated matrix of the operator $O$ with
respect to the same orthonormal basis (\ref{entI}). Notice that for the
particular state in Eq. (\ref{rhoN}), one has $\rho_{\cal N}^{\mathrm
t}\equiv \rho_{\cal N}$, since $\ket{\psi_0}^{\otimes N}$ has all
real coefficients on the basis (\ref{entI}).
The covariance (\ref{phicov}) of the CP map $\Cmap_{NM}$ in terms of the operator
$R_{NM}$ becomes the invariance relation
\begin{equation}
[R_{NM},U_\phi^{\otimes M}\otimes (U_\phi^{\otimes N})^*]=0,
\end{equation}
and in our case we have simply
$(U_\phi^{\otimes N})^*\equiv U_{-\phi}^{\otimes N}$. Then, according to the
Schur lemmas the positive operator $R_{NM}$ is given by the following
direct sum
\begin{equation}
R_{NM}=\oplus_\nu R_\nu,
\end{equation}
where $\nu$ runs over all inequivalent unitary irreducible
representations (UIR) contained in the reducible one $U_\phi^{\otimes
M}\otimes (U_\phi^{\otimes N})^*$, with all equivalent representations
grouped together, and with $R_\nu$ denoting any positive operator
over the space of all representations equivalent to $\nu$, with the
overall constraint of the trace-preserving condition (\ref{trcond}).
\par Our purpose is to find the optimal phase-covariant cloning map
which maximizes the fidelity in Eq. (\ref{fidglob3}), which using Eq. (\ref{CR})
can be rewritten in terms of the positive operator $R_{NM}$ as follows 
\begin{equation}
f_{NM}=\Tr[(\ket{\psi_0}\bra{\psi_0}^{\otimes M}\otimes
\ket{\psi_0}\bra{\psi_0}^{\otimes N}) R_{NM}].
\end{equation}
The derivation for the case of qutrits will be strictly analogous to
that of qubits.

\section{Optimal cloning for qubits}

Since the phase rotation group is abelian, all UIR's of the group are
unidimensional. The inequivalent representations can be conveniently
labeled by the nonnegative integer $\nu$, corresponding to the invariant
spaces of vectors where the group action is equivalent to
multiplication by the phase factor $\exp(i\nu\phi)$. Therefore, in the
reduction of the representation $U_\phi^{\otimes
M}\otimes U_{-\phi}^{\otimes N}$, each UIR equivalent to the
representation $\nu$ is spanned by a vector of 
the type
\begin{equation}
\begin{split}
&\ket{M-j-\nu,j+\nu}\otimes\ket{N-j,j},\\ & j=0,\ldots,
\min(N,M-\nu),\quad\nu=0,\ldots, M-N,
\end{split}
\end{equation}
where $\ket{N-j,j}$
denotes a state of $N$ qubits where $N-j$ of them are in state $\ket{0}$, while
the remaining $j$ are in state $\ket{1}$.

We will now look for the optimal transformations, namely the transformations
that maximize the fidelity $f_{NM}$.
As proved above, we can restrict our attention 
to the symmetric subspace and therefore
we will consider the equivalent representations corresponding  to the 
symmetric states
$\{\ket{s_{M,j+\nu}}\ket{s_{N,j}},j=0,...\min(N,M-\nu)\}_\nu$, 
where $\nu$ labels the 
inequivalent representations ($\nu=0,M-N$).
In the evaluation of the fidelity we take 
$\ket{\psi_0}=(\ket{0}+\ket{1})/\sqrt{2}$. 
The fidelity of the map is made of contributions of the form
\begin{equation}
\begin{split}
& {\mbox{Tr}}[(\proj{\psi_0}^{\otimes M+N})
(\ket{s_{M,j+\nu}}\bra{s_{M,k+\nu}}\otimes 
\ket{s_{N,j}}\bra{s_{N,k}})]\\
&=\tfrac{1}{2^{N+M}}\sqrt{C(N,j)C(N,k)}\sqrt{C(M,j+\nu)C(M,k+\nu)}\;.
\end{split}
\label{contr}
\end{equation}
Each block of equivalent representations labeled by $\nu$ is given by
the positive operator
\begin{equation}
R_\nu=\sum_{jk}r_{jk}^\nu\ket{s_{M,j+\nu}}\bra{s_{M,k+\nu}}\otimes\ket{s_{N,j}}
\bra{s_{N,k}},
\end{equation}
where the trace preserving condition for the operator $R_{NM}$ leads to
\begin{equation}
\sum_{\nu=0}^{M-1} r_{jj}^\nu=1\;,\quad i=0,...N\;.
\label{cr}
\end{equation}
Since each single contribution to the fidelity (\ref{contr}) is
positive versus $j$ and $k$, the operators $R_\nu$ that maximize the
fidelity have positive elements $r_{jk}^\nu$
and the off diagonal terms are as large as possible, i.e. $r_{jk}^\nu=
\sqrt{r_{jj}^\nu}\sqrt{r_{kk}^\nu}$. Therefore, the operator $R_\nu$ can
be written as a (generally non normalized) projector $R_\nu=\proj{r_\nu}$,
where $\ket{r_\nu}=\sum_j r_j^\nu\ket{s_{M,j+\nu}}\otimes\ket{s_{N,j}}$,
and $r_j^\nu\doteq\sqrt{r_{jj}^\nu}$.

Let us now explicitly construct the cloning map that optimizes the fidelity.
We will first consider the simple case $N=1$. 
Each term $R_\nu$ will therefore give the following contribution to $f_{1M}$
\begin{equation}
\begin{split}
f_{1M}^\nu&=\Tr[(\ket{\psi_0}\bra{\psi_0}^{\otimes M}\otimes
\ket{\psi_0}\bra{\psi_0}) R_{\nu}]\\&=
\tfrac{1}{2^{M+1}}\left(r_0^\nu\sqrt{C(M,\nu)}+r_1^\nu\sqrt{C(M,1+\nu)}
\right)^2
\label{fnu1M}
\end{split}
\end{equation}
with $f_{1M}=\sum_\nu f_{1M}^\nu$. 

For odd values of $M$ the largest contribution to the fidelity comes from 
the projector with $\bar\nu=(M-1)/2$ because in this case both terms
$\sqrt{C(M,\nu)}$ and $\sqrt{C(M,1+\nu)}$ are equal and are 
maximized simultaneously. 
Moreover, this contribution is maximized when 
the values of $r_0$ and $r_1$ are maximized, i.e. for 
$r_0^{(M-1)/2}=r_1^{(M-1)/2}=1$.
In this case the optimal map is given by
\begin{equation}
R_{1M}=\ket{r_{(M-1)/2}}\bra{r_{(M-1)/2}}
\label{op1m}
\end{equation}
and the fidelity takes the form 
\begin{equation}
f_{1M}=\tfrac{1}{2^{M-1}}C(M,(M-1)/2)\;.
\end{equation}

For even values of $M$ the optimization procedure is more involved because 
the coefficients
$\sqrt{C(M,\nu)}$ and $\sqrt{C(M,1+\nu)}$ are different and 
cannot be maximized simultaneously by a single value of $\nu$.
In order to derive the form of the optimal map
let us first notice that  the same contribution $f_{1M}^\nu$ in 
Eq. (\ref{fnu1M}) is also achieved
by choosing $\nu'=M-\nu-1$ with $r_0^\nu=r_1^{M-\nu-1}$ and 
\begin{equation}
r_1^\nu=r_0^{M-\nu-1}\;.
\label{eqtwo} 
\end{equation}
Therefore we can look at contributions due to maps of the form
\begin{equation}
R'_{\nu}=\tfrac{1}{2}(R_\nu+R_{M-\nu-1}).
\label{twor}
\end{equation}
By taking into account the relations (\ref{eqtwo}), in this case the 
completeness constraint (\ref{trcond}) can be written as
\begin{equation}
\sum_{\nu}(r_0^{\nu})^2+
\sum_{\nu}(r_1^{\nu})^2=2\;.
\label{k}
\end{equation}
The optimal map is given by the values of $\nu$ that give the maximum 
contributions to (\ref{fnu1M}), namely for $\nu_-=M/2-1$ and 
$\nu_+=M/2$. Therefore the optimization problem consists in maximizing the 
quantity $r_0^{\nu_-}A+r_1^{\nu_-}B$, with the constraint
$(r_0^{\nu_-})^2+(r_1^{\nu_-})^2=2$ and with $A=\sqrt{C(M,\nu_-)}$
and $B=\sqrt{C(M,\nu_-+1)}$.
The solution is given by
\begin{equation}
r_0^{\nu_-}=\sqrt{2}\tfrac{B}{\sqrt{A^2+B^2}}\;,\qquad 
r_1^{\nu_-}=\sqrt{2}\tfrac{A}{\sqrt{A^2+B^2}}\;.
\label{sol}
\end{equation}
Therefore, the optimal map $R_{1M}$ for even values of $M$ is given by
\begin{equation}
R_{1M}=\tfrac{1}{2}(\ket{r_{\nu_-}}\bra{r_{\nu_-}}+
\ket{r_{\nu_+}}\bra{r_{\nu_+}})\;,
\end{equation}
with $r_0^{\nu_-}$ and $r_1^{\nu_+}$ given by Eq. (\ref{sol}), and 
$r_0^{\nu_+}=r_1^{\nu_-}$, $r_1^{\nu_+}=r_0^{\nu_-}$.  
The fidelity takes the form
\begin{equation}
f_{1M}=\tfrac{1}{2^{M-1}}\tfrac{C(M,M/2-1)C(M,M/2)}{[C(M,M/2-1)^2+C(M,M/2)]^2}\;.
\end{equation}
Consider now the general case $N\to M$. 
Each contribution $f_{NM}^\nu$ to the fidelity takes the form
\begin{equation}
f_{NM}^\nu=\tfrac{1}{2^{N+M}}\left[\sum_{j=0}^{\min(N,M-\nu)} 
\!\!\!\!\!\!\! r_j^\nu \sqrt{C(N,j)C(M,j+\nu)}\right]^2\!\!\!\!.\!\!\!\label{contfnm}
\end{equation}
The maximum value is achieved for the representation $\bar\nu$ for
which both the terms $\sqrt{C(N,j)}$ and 
$\sqrt{C(M,j+\nu)}$ are maximized at the  same time.
In fact, $\sum_\nu f_{NM}^\nu$ is a convex function of $r_j^\nu$ defined 
on the convex domain (\ref{cr}), and the maximum is achieved on the extremal 
points $r_{jj}^\nu=1$ for some $\nu$. This also corresponds to maximize
the r.h.s. of Eq. (\ref{contfnm}) by adding ``coherently'' all the terms in the
sum over $j$ for a single value of $\nu$.
We have then to distinguish different cases: for $N$ odd and $M$ odd
the simultaneous maximization of $\sqrt{C(N,j)}$ and $\sqrt{C(M,j+\nu)}$ 
occurs when $j=(N-1)/2$ and $\bar\nu=(M-N)/2$. In this case the optimal 
cloning map corresponds to $r_{j}^{\bar\nu}=1$, and is described by
\begin{equation}
R_{NM}=\proj{r_{(M-N)/2}}\;.
\label{opt_rnm}
\end{equation}
The fidelity takes the explicit form
\begin{equation}\!
f_{NM}=\tfrac{1}{2^{N+M}}\!\!\!\left[\sum_{j=0}^N \sqrt{C(N,j)C(M,(M-N)/2+j)}
\right]^2\!\!\!\!.\!\!\!\!\!\!\!
\label{opt_fnm}
\end{equation}
An analogous argument and the results given in Eqs. (\ref{opt_rnm}) and 
(\ref{opt_fnm}) hold also when $M$ and $N$ are both even.

Consider now the case of even $M$ and odd $N$, or viceversa. 
The two terms $\sqrt{C(N,j)}$ and $\sqrt{C(M,j+\nu)}$
are maximized at the same time for the two values $\nu_\pm=(M-N\pm 1)/2$.
In order to derive the optimal map we follow an argument analogous to the
case $N=1$ discussed above. 
Actually, let us first notice that the same contribution $f_{NM}^\nu$ in 
Eq. (\ref{contfnm}) is also achieved by choosing $\nu'=M-N-\nu$ 
with $r_j^\nu=r_{N-j}^{M-N-\nu}$.  
Therefore, as in the case $N=1$ we can look at cloning maps of the form
\begin{equation}
R'_{\nu}=\tfrac{1}{2}(R_\nu+R_{M-N-\nu})\;.
\label{twornm}
\end{equation}
By exploiting the relation $r_j^\nu=r_{N-j}^{M-N-\nu}$, we can write the 
completeness condition as 
\begin{equation}
\sum_{\nu}(r_j^{\nu})^2+
\sum_{\nu}(r_{N-j}^{\nu})^2=2\;,\qquad j=0,..N/2.
\label{k2}
\end{equation}
As mentioned above, the greatest contributions to ther fidelity are given by 
the blocks with $\nu_\pm=(M-N\pm 1)/2$. The optimal cloning map will therefore
be of the form
\begin{equation}
R_{NM}=\tfrac{1}{2}(R_{\nu_-}+R_{\nu_+})\;,
\label{twornm2}
\end{equation}
with the constraints $r_j^{\nu_+}=r_{N-j}^{M-N-\nu_-}$ and
$(r_j^{\nu_-})^2+(r_{N-j}^{\nu_-})^2=2\;,j=0,..N/2$.
The optimization of the fidelity (\ref{contfnm}) with the constraints 
(\ref{k2}) leads to the following solutions
\begin{equation}
\begin{split}
r_j^{\nu_-}&=\sqrt{2}\tfrac{\sqrt{C(M,N-j+\nu_-)}}{\sqrt{C(M,j+\nu_-)
+C(M,N-j+\nu_-)}},\\
r_{N-j}^{\nu_-}&=\sqrt{2}\tfrac{\sqrt{C(M,j+\nu_-)}}{\sqrt{C(M,j+\nu_-)
+C(M,N-j+\nu_-)}}\;.
\label{sol2}
\end{split}
\end{equation}
Let us now consider as a quality criterion to optimize the cloning map
the optimization of the average 
single particle fidelity $F_{NM}$, defined as
\begin{equation}
\begin{split}
&F_{NM}=\tfrac{1}{M}{\mbox{Tr}}[(\proj{\psi_0}\eins^{\otimes M-1}
+\eins \proj{\psi_0}\eins^{\otimes M-2}+...\\
&+\eins^{\otimes M-1} \proj{\psi_0})\otimes\proj{\psi_0}^{\otimes N}R_{NM}].\label{F}
\end{split}
\end{equation}
In this case we assume that the operators $R_{NM}$ are supported on the 
symmetric 
subspace $({\cal{H}}^{\otimes M})_+$. Notice that the last requirement is now
an assumption because the argument after Eq. (\ref{fidglob3}), valid
for the global fidelity $f_{NM}$, does not hold for the average single 
particle fidelity. In this case the fidelity of the map is made of 
contributions of the form
\begin{equation}
\begin{split}
&{\mbox{Tr}}[(\proj{\psi_0}\eins^{\otimes M-1}\proj{\psi_0}^{\otimes
  N})\\ &\times (\ket{s_{M-j-\nu}}\bra{s_{M-k-\nu}}\otimes 
\ket{s_{N-j}}\bra{s_{N-k}})]\\
=&\tfrac{1}{2^{N+1}}\left[C(N,j)\delta_{j,k}+
\tfrac{\sqrt{C(N,j)C(N,j+1)}C(M-1,j+\nu)}{\sqrt{C(M,j+\nu)C(M,j+\nu+1)}}
\delta_{j+1,k}\right]\\
=&\tfrac{1}{2^{N+1}}\left[C(N,j)\delta_{j,k}+\tfrac{1}{M}\sqrt{C(N,j)C(N,j+1)}
\right.
\\ &\left.\times \sqrt{(M-j-\nu)(j+\nu+1)}\delta_{j+1,k}\right]\;,
\end{split}
\end{equation}
where we have considered $k\geq j$. 
As in the case of the global fidelity $f_{NM}$, let us start 
from the case $N=1$.

Each term $R_\nu$ will therefore give the following contribution to $F_{1M}$
\begin{equation}
F_{1M}^\nu=
\tfrac{1}{4}\left[(r_0^\nu)^2+(r_1^\nu)^2+\tfrac{2}{M}r_0^\nu r_1^\nu
\sqrt{(M-\nu)(\nu+1)}\right]\;.
\label{fnu1m}
\end{equation}
For odd values of $M$ the term $\sqrt{(M-\nu)(\nu+1)}$ is maximized for
$\nu=(M-1)/2$. The optimal map, as in the case of the optimization 
of the global fidelity, is given by Eq. (\ref{op1m}) with 
$r_0^{(M-1)/2}=r_1^{(M-1)/2}=1$. The fidelity in this case takes the explicit 
form
\begin{equation}
F_{1M}=\tfrac{1}{2}\left(1+\tfrac{M+1}{2M}\right)\;.
\end{equation}
For odd values of $M$, we can argue similarly to the case of the global 
fidelity, and therefore we have to maximize the quantity (\ref{fnu1m}) with the
constraint $(r_0^{\nu_-})^2+(r_1^{\nu_-})^2=2$. In this case the optimal
solution corresponds to $r_0^{\nu_-}=r_1^{\nu_-}=1$. The form of the optimal 
map is given by
\begin{equation}
R_{1M}=\lambda \ket{r_{\nu_-}}\bra{r_{\nu_-}}+
(1-\lambda) \ket{r_{\nu_+}}\bra{r_{\nu_+}}\;,
\end{equation}
with $0\leq \lambda \leq 1$,
and the fidelity takes the form
\begin{equation}
F_{1M}=\tfrac{1}{2}\left(1+\tfrac{\sqrt{M(M+2)}}{2M}\right)\;.
\end{equation}
The above optimal single particle fidelities are the same as the ones 
reported in Ref. 
\cite{cinesi}, where cloning transformations restricted to the symmetric 
subspace were studied and the optimality of the single particle
fidelity was proved only for $N=1$.

Consider now the general case $N\to M$. 
Each contribution $F_{NM}^\nu$ to the fidelity takes the form
\begin{equation}
\begin{split}
F_{NM}^\nu&=\tfrac{1}{2^{N+1}}\left[\sum_{j=0}^{\min(N,M-\nu)}  
(r_{j}^{\nu})^2 C(N,j) \right.\\ &
+\tfrac{2}{M}\sum_{j=0}^{\min(N,M-\nu)-1}r_j^\nu r_{j+1}^\nu
\sqrt{C(N,j)C(N,j+1)}\\ &\times\left.
\sqrt{(M-j-\nu)(j+\nu+1)}\right].\label{Fnu}
\end{split}
\end{equation}
The maximum value is achieved for the representation $\nu$ for
which both the terms $\sqrt{C(N,j)C(N,j+1)}$ and 
$\sqrt{(M-j-\nu)(j+\nu+1)}$ are maximized at the  same time. 
We have then to distinguish different cases: for $N$ odd and $M$ odd
this occurs when $j=(N-1)/2$ and $\bar\nu=(M-N)/2$. In this case the optimal 
cloning map corresponds to $r_{j}^{\bar\nu}=1$, and is described by
\begin{equation}
R_{NM}=\proj{r_{(M-N)/2}}\;.\label{opt_rnm2}
\end{equation}
The fidelity in this case takes the explicit form
\begin{equation}
\begin{split}
F_{NM}&=\tfrac{1}{2}
+\tfrac{1}{M 2^N}\sum_{j=0}^{N-1}\sqrt{C(N,j)C(N,j+1)}\\
&\times\sqrt{[(M+N)/2-j][(M-N)/2+j+1]}.\label{opt_fnm2}
\end{split}
\end{equation}
The results given in Eqs. (\ref{opt_rnm2}) and (\ref{opt_fnm2}) hold also when 
$M$ and $N$ are both even. Notice that these results are in agreement with 
the ones conjectured in Ref. \cite{cinesi} for generic $N$ and $M$.

Consider now the case where $N$ and $M$ have different parity, for example 
$M$ is even and $N$ is odd.
The two terms $\sqrt{C(N,j)C(N,j+1)}$ and $\sqrt{(M-j-\nu)(j+\nu+1)}$
are maximized at the same time for the two values $\nu_\pm=(M-N\pm 1)/2$.
The optimal cloning map will therefore be of the form
\begin{equation}
R_{NM}=\tfrac{1}{2}(R_{\nu_-}+R_{\nu_+})\;,
\label{twornm3}
\end{equation}
with the constraints $r_j^{\nu_+}=r_{N-j}^{\nu_-}$ and
$(r_j^{\nu_-})^2+(r_{N-j}^{\nu_-})^2=2\;,j=0,..N/2-1$.
Therefore, we have to optimize the fidelity (\ref{Fnu}) with $\nu=\nu_-$
by taking into account the above constraints, namely the quantity
\begin{equation}
\begin{split}
F_{NM}^{\nu_-}=&\tfrac{1}{2}+\tfrac{1}{2^{N}M}
\sum_{j=0}^{N-1}r_j^{\nu_-} r_{j+1}^{\nu_-}
\sqrt{C(N,j)C(N,j+1)}\\ \times&\sqrt{(M-j-\nu_-)(j+\nu_-+1)}\;.
\label{Fnu-}
\end{split}
\end{equation}
The forms of the coefficients $r_j$ cannot be found in general.
As an example we explicitly optimize the fidelity for $N=2$ and odd $M$. 
In this case $\nu_-=(M-3)/2$ and the form of the coefficients is given by 
\begin{equation}
\begin{split}
r_0^{\nu_-}=&\sqrt{2}\tfrac{\sqrt{(M-1)(M+3)}}{\sqrt{(M-1)(M+3)+(M+1)^2}},\\
r_1^{\nu_-}=&1 \;,\quad r_2^{\nu_-}=\sqrt{2-(r_0^{\nu_-})^2}\;,
\label{r2m}
\end{split}
\end{equation}
and the fidelity takes the explicit form
\begin{equation}
F_{2M}=\tfrac{1}{2}\left(1+\tfrac{\sqrt{M^2+2M-1}}{\sqrt{2}M}\right)\;.
\end{equation}
We can see that in general when $N$ and $M$ have different parity
the optimal solutions are not in agreement with the
optimal transformations conjectured in \cite{cinesi}.

We want to point out that the fidelity $F_{NM}$ of the above optimal 
cloning transformations in the limit $M\to\infty$ coincides with the 
fidelity of optimal state estimation for $N$ equatorial qubits \cite{dbe}.

Moreover, we want to stress that 
the cloning transformations that optimize the global fidelity
coincide with the optimal ones for the single particle fidelity only in
the cases where $N$ and $M$ have the same parity.
 
\section{Optimal cloning for qutrits}
In this section we will derive the optimal $1\to M$ cloning transformations for
equatorial qutrit states
\begin{equation}
\ket{\psi_{\phi,\theta}}=\tfrac{1}{\sqrt 3}(\ket{0}+e^{i\phi}\ket{1}
+e^{i\theta}\ket{2}),
\label{def-qutrit}
\end{equation}
covariant under the group of rotations of both phases $\phi$
and $\theta$. Again, since the group is abelian, all UIR's of the
group are unidimensional, and in a way analogous to the case of
cloning of qubits, when restricting to output states supported on the
symmetric subspace $(\sH^{\otimes N})_+$, the equivalent
UIR's are spanned by the vectors
\begin{equation}
\ket{s_{M,\nu_1,\nu_2}}\ket{0},\quad
\ket{s_{M,\nu_1+1,\nu_2}}\ket{1},\quad
\ket{s_{M,\nu_1,\nu_2+1}}\ket{2}
\end{equation}
where $\nu_1=0,\ldots,M-1$ and $\nu_2=0,\ldots,M-\nu_1-1$
label the invariant spaces of the UIR's corresponding to
multiplication by the phase factor $\exp(i\nu_1\phi+i\nu_2\theta)$,
and $\ket{s_{k,p,q}}$ denotes the 
normalized symmetric state of $k$ qutrits with $k-p-q$ qutrits in state 
$\ket{0}$, $p$ in state $\ket{1}$ and $q$ in state $\ket{2}$
(the state $\ket{s_{k,p,q}}$ is a superposition of $k!/(k-q-p)!p!q!$ orthogonal
states). 

In this case we will have contributions of the following type to the fidelity 
\begin{equation}
\begin{split}
&{\mbox{Tr}}[\proj{\psi_{0,0}}^{\otimes M+1}
(\ket{s_{M,\nu_1,\nu_2}}\bra{s_{M,\nu_1+1,\nu_2}}\otimes 
\ket{0}\bra{1})]\\
&=\tfrac{1}{3^{M+1}}\sqrt{T(M,\nu_1,\nu_2)}\sqrt{T(M,\nu_1+1,\nu_2)}
\end{split}
\end{equation}
where we define 
\begin{equation}
T(M,\nu_1,\nu_2)=\tfrac{M!}{(M-\nu_1-\nu_2)!\nu_1!\nu_2!}\;.
\end{equation}
Since all the above contributions are positive, we can apply the same
argument as in the case of qubits and consider positive operators of the form
$R_{\nu_1,\nu_2}=\proj{r_{\nu_1,\nu_2}}$,
where 
\begin{equation}
\begin{split}
&\ket{r_{\nu_1,\nu_2}}=r_0^{\nu_1,\nu_2}
\ket{s_{M,\nu_1,\nu_2}}\ket{0}\\&+
r_1^{\nu_1,\nu_2}\ket{s_{M,\nu_1+1,\nu_2}}\ket{1}
+r_2^{\nu_1,\nu_2}\ket{s_{M,\nu_1,\nu_2+1}}\ket{2},
\end{split}
\label{ketr}
\end{equation}
and the trace preserving condition for the operator $R_{1M}$ leads to
\begin{equation}
\sum_{\nu_1,\nu_2}(r_0^{\nu_1,\nu_2})^2=
\sum_{\nu_1,\nu_2}(r_1^{\nu_1,\nu_2})^2
=\sum_{\nu_1,\nu_2}(r_2^{\nu_1,\nu_2})^2=1\;,
\label{trp}
\end{equation}
where in each sum $\nu_1$ and $\nu_2$ are constrained to give non negative
entries in the states in Eq. (\ref{ketr}).

Each operator $R_{\nu_1,\nu_2}$ gives the following contribution to the 
fidelity
\begin{equation}
\begin{split}
&f_{\nu_1,\nu_2}=\tfrac{1}{3^{M+1}}\left(r_0^{\nu_1,\nu_2}
\sqrt{T(M,\nu_1,\nu_2)}\right.\\&\left. +r_1^{\nu_1,\nu_2}\sqrt{T(M,\nu_1+1,\nu_2)}+
r_2^{\nu_1,\nu_2}\sqrt{T(M,\nu_1,\nu_2+1)}\right)^2.
\label{fnu1nu2}
\end{split}
\end{equation}
The operator $R_{\nu_1,\nu_2}$ that gives the highest contribution to the
fidelity is the one where the values of $T(M,\nu_1,\nu_2)$, 
$T(M,\nu_1+1,\nu_2)$ and $T(M,\nu_1,\nu_2+1)$ are maximized. 
This is easy to establish in the case of $M=3k+1$, because the three above 
expressions for $T$ are all simultaneously maximized for $\nu_1=\nu_2=k$. 
Therefore, the optimal cloning map is given by
$R_{k,k}$ with $r_0=r_1=r_2=1$. 
The corresponding fidelity takes the explicit form
\begin{equation}
f_{M}=\tfrac{1}{3^{M-1}}T\left(M,\tfrac{M-1}{3},\tfrac{M-1}{3}\right)\;.
\end{equation}

The cases with $M=3k$ and $M=3k+2$ are more involved because the
three values of  
$T$ that appear in Eq. (\ref{fnu1nu2}) cannot be maximized simultaneously.
In order to find the form of the optimal maps we will follow an argument
similar to the case of qubits.
Notice first that the value of the contribution $f_{\nu_1,\nu_2}$ to the 
fidelity does not change by
performing any permutation of the basis states $\{\ket{0},\ket{1},\ket{2}\}$ 
for each of the $M+1$ qutrits in the operator $R_{\nu_1,\nu_2}$.
This means that the three blocks labelled by $(\nu_1,\nu_2)$, 
$(\nu_2,M-\nu_1-\nu_2-1)$ and $(M-\nu_1-\nu_2-1,\nu_1)$ give the same 
contribution  to the fidelity. Therefore, the same contribution given by
the operator $R_{\nu_1,\nu_2}$ is achieved also by the map 
\begin{equation}
R'_{1M}=\tfrac{1}{3}(R_{\nu_1,\nu_2}+R_{\nu_2,M-\nu_1-\nu_2-1}
+R_{M-\nu_1-\nu_2-1,\nu_1})\;,
\end{equation}
with the following identifications
\begin{equation}
\begin{split}
r_0^{\nu_1,\nu_2}&=r_1^{\nu_2,M-\nu_1-\nu_2-1}=r_2^{M-\nu_1-\nu_2-1,\nu_1}\;,
\\ 
r_1^{\nu_1,\nu_2}&=r_2^{\nu_2,M-\nu_1-\nu_2-1}=r_0^{M-\nu_1-\nu_2-1,\nu_1}\;,
\\ 
r_2^{\nu_1,\nu_2}&=r_0^{\nu_2,M-\nu_1-\nu_2-1}=r_1^{M-\nu_1-\nu_2-1,\nu_1}\;.
\end{split}
\label{rperm}
\end{equation}
The completeness constraint (\ref{trcond}) along with Eqs. (\ref{rperm}) lead 
to
\begin{equation}
\sum_{\nu_1,\nu_2}(r_0^{\nu_1,\nu_2})^2+
\sum_{\nu_1,\nu_2}(r_1^{\nu_1,\nu_2})^2+
\sum_{\nu_1,\nu_2}(r_2^{\nu_1,\nu_2})^2=3\;.
\label{norm2}
\end{equation}
If we restrict our attention to the family of cloning transformations 
described by $R_{\nu_1,\nu_2}$ we have to fulfill the constraint 
\begin{equation}
(r_0^{\nu_1,\nu_2})^2+(r_1^{\nu_1,\nu_2})^2+(r_2^{\nu_1,\nu_2})^2=3\;.
\label{normR}
\end{equation}
Let us first consider the case $M=3k$.
From Eq. (\ref{fnu1nu2}) we can see that the 
representation that contributes mostly to the fidelity is the one with
$\bar\nu_1=\bar\nu_2=k$, because one of the three coefficients $T$ that appear
in (\ref{fnu1nu2}) is maximized
and the other two take the second possible highest value simultaneously. 
Therefore, we can maximize the fidelity by restricting our attention to
the block labelled by $\bar\nu_1$ and $\bar\nu_2$.  
Moreover, since $T(M,\nu_1+1,\nu_2)$ and $T(M,\nu_1,\nu_2+1)$ 
have the same value for $\nu_1=\nu_2=k$, the expression (\ref{fnu1nu2})
is invariant under exchange of the coefficients $r_1^{k,k}$ and $r_2^{k,k}$.  
Therefore we can set $r_1^{k,k}=r_2^{k,k}$ when we look for the optimal 
solution.
The optimization of the contribution in Eq. (\ref{fnu1nu2}) with 
$\nu_1=\nu_2=k$ corresponds 
to maximizing the quantity $r_0^{k,k}A + 2r_1^{k,k}B$, with the constraint 
$(r_0^{k,k})^2+2(r_1^{k,k})^2=3$ and with $A=\sqrt{T(M,M/3,M/3)}$ and 
$B=\sqrt{T(M,M/3+1,M/3)}$. The solution corresponds to 
\begin{equation}
\begin{split}
r_0^{k,k}=&\sqrt{3-2(r_1^{k,k})^2},\\
r_1^{k,k}=&\sqrt{3}\left(\tfrac{A^2}{B^2}+2\right)^{-\tfrac{1}{2}},
\quad r_2^{k,k}=r_1^{k,k},
\end{split}
\label{optr0r1}
\end{equation}
and all other nonvanishing coefficients given by Eq. (\ref{rperm}).
The corresponding optimal map is then given by
\begin{equation}
R_{1M}=\tfrac{1}{3}(R_{M/3,M/3}+R_{M/3,M/3-1}+R_{M/3-1,M/3})\;,
\label{optrqutrits}
\end{equation}
 with the nonvanishing coefficients $r_i^{\nu_1,\nu_2}$ 
given by Eqs. (\ref{rperm}) and (\ref{optr0r1}).

In the remaining case $M=3k+2$ the optimization argument and the final 
solution are the same as for $M=3k$. Here we maximize the quantity
$r_0^{k,k} A+2r_1^{k,k}B$, with $A=T(M,(M-2)/3,(M-2)/3)$ and 
$B=T(M,(M-2)/3,(M-2)/3+1)$. The optimal solution can be derived analogously 
to the previous case.

As in the case of qubits we can derive optimal maps for qutrits by maximizing
the average single particle fidelity instead of the global one, 
and by assuming that the operator $R$ is supported on the symmetric subspace. 
As in the case of qubits we will see that the optimal
maps for the average single particle fidelity are not always the same
as the ones derived above, where the global fidelity was maximized.
Actually, in this case the contributions to the average single particle 
fidelity $F_{1M}$ are of the form
\begin{equation}
\begin{split}
&\tfrac{1}{9}{\mbox{Tr}}[(\proj{\psi_{0,0}}
\eins^{\otimes M-1}\proj{\psi_{0,0}})\\ &\times
\ket{s_{M,\nu_1,\nu_2}}\bra{s_{M,\nu_1+1,\nu_2}}\otimes 
\ket{0}\bra{1}]\\
&=\tfrac{1}{9}\tfrac{(M-1)!}{(M-\nu_1-\nu_2-1)!\nu_1!\nu_2!}
\sqrt{\tfrac{(M-\nu_1-\nu_2)!\nu_1!\nu_2!}{M!}}\\ &\times
\sqrt{\tfrac{(M-\nu_1-\nu_2-1)!(\nu_1+1)!\nu_2!}{M!}}\\
&=\tfrac{1}{9M}\sqrt{(M-\nu_1-\nu_2)(\nu_1+1)}\\ &
\equiv\tfrac{1}{9}\Lambda_M(M-\nu_1-\nu_2,\nu_1)\;,
\end{split}
\end{equation}
where we define
\begin{equation}
\begin{split}
\Lambda_M(p,q)&\equiv{\mbox{Tr}}[(\proj{\psi_{0,0}}
\eins^{\otimes M-1})\ket{s_{M,p,q}}\bra{s_{M,p-1,q+1}}]\\
&=\tfrac{1}{M}\sqrt{p(q+1)}.
\end{split}
\end{equation}
The arguments leading to the form (\ref{ketr}) and to the constraints  
(\ref{norm2}) hold also in this case.
The contributions to the fidelity due to the operators $R_{\nu_1,\nu_2}$ 
are given by
\begin{equation}
\begin{split}
F_{\nu_1,\nu_2}&=\tfrac{1}{9}[(r_0^{\nu_1,\nu_2})^2+(r_1^{\nu_1,\nu_2})^2+
(r_2^{\nu_1,\nu_2})^2\\
&+2r_0^{\nu_1,\nu_2}r_1^{\nu_1,\nu_2}\Lambda_M(M-\nu_1-\nu_2,\nu_1)\\ &
+2r_0^{\nu_1,\nu_2}r_2^{\nu_1,\nu_2}\Lambda_M(M-\nu_1-\nu_2,\nu_2)\\ &
+2r_1^{\nu_1,\nu_2}r_2^{\nu_1,\nu_2}\Lambda_M(\nu_1+1,\nu_2)]\;.
\label{Fnu1nu2}
\end{split}
\end{equation}
As discussed above, the optimal cloning map corresponds to optimizing the 
coefficients $r_i$ for the block $R_{\nu_1,\nu_2}$ that gives the maximum 
contribution (\ref{Fnu1nu2}). In the case $M=3k+1$, all the three terms 
$\Lambda_M$ 
that appear in the expression  (\ref{Fnu1nu2}) are optimized at the same time 
for $\nu_1=\nu_2=k$, and therefore the optimal map has the same form as the
one found by maximizing the global fidelity. The fidelity $F_{1M}$ in this case
takes the explicit form
\begin{equation}
F_{M}=\tfrac{1}{3}\left(1+2\tfrac{M+2}{3M}\right)\;.
\end{equation}
Let us now consider the case of $M=3k$.
By looking at Eq. (\ref{Fnu1nu2}) we can see that the 
maximum contribution to the fidelity corresponds to
$\nu_1=\nu_2=k$, because one of the three coefficients $\Lambda_M$ is optimized
and the other two take the second possible highest value simultaneously. 
Moreover, since $\Lambda_M(M-\nu_1-\nu_2,\nu_1)$ and 
$\Lambda_M(M-\nu_1-\nu_2,\nu_2)$
have the same value for $\nu_1=\nu_2=k$, the expression (\ref{Fnu1nu2})
the optimal solution corresponds to $r_1^{k,k}=r_2^{k,k}$.  
Therefore, the maximum contribution (\ref{Fnu1nu2}) corresponds to 
maximizing the quantity
\begin{equation}
2r_0^{k,k}r_1^{k,k} \Lambda_A + (r_1^{k,k})^2 \Lambda_B\;,
\label{op}
\end{equation}
with the constraint $(r_0^{k,k})^2+2(r_1^{k,k})^2=3$ and with 
$\Lambda_A=\sqrt{M(M+3)}/3M $ and 
$\Lambda_B=(M+3)/3M$. The optimal solution corresponds to 
\begin{equation}
\begin{split}
r_0^{k,k}&=\sqrt{3-2(r_1^{k,k})^2},\\ 
r_1^{k,k}&=\tfrac{\sqrt 3}{2}\sqrt{1+\tfrac{\Lambda_B}{\sqrt{\Lambda_B^2+
8\Lambda_A^2}}},\qquad 
r_2^{k,k}=r_1^{k,k}\;.
\label{b_op}
\end{split}
\end{equation}
The optimal map has the form (\ref{optrqutrits}),
with the values of the coefficients $r_i$ for eack block fixed according to 
Eqs. (\ref{rperm}) and (\ref{b_op}).

In the remaining case $M=3k+2$ the optimization argument and the final 
solution are the same as in the $M=3k$ case, with 
$\Lambda_A=\sqrt{(M+4)(M+1)}/3M$ and $\Lambda_B=(M+1)/3M$. Notice that the 
optimal map coincide with the case $1\to 3$ derived in \cite{dalop}.
 
We want to point out that the average single particle fidelity in the limit
$M\to\infty$ coincides with the fidelity of optimal double-phase estimation
for a qutrit in the state (\ref{def-qutrit}) \cite{ph}.

\section{Discussion}
In this paper we derived from first principles the optimal quantum
cloning transformations that maximize the fidelity averaged uniformly over
all equatorial qubit states. We have seen that such averaged form for
the fidelity automatically leads to the optimal phase-covariant
cloning. We have then derived the optimal $N\to M$ cloning transformation 
using the method \cite{dalop}
designed for group-covariant cloning. We have also considered
phase-covariant cloning for qutrits, and derived the $1\to M$
optimal cloning maps. From our analytical results one
can see that the fidelities are always larger than those obtained for
the universal cloning \cite{Werner}. Moreover, the fidelity
for the qutrit cloning is smaller than the corresponding
one for the qubit. We also found that the form of the optimal cloning maps 
depends on the criterion adopted to assess the quality of the transformation.
Actually, we showed that the maximization of the global fidelity and the
maximization of the average single particle fidelity in general lead to
different solutions.

\par We want now to emphasize that the general analysis performed in Sect. II for
optimal phase-covariant cloning would be exactly the same for any smaller
discrete phase-covariance group, such as for example the discrete group
${\mathbf Z}_4$ of $\pi/2$-rotations that is 
employed in the BB84 cryptographic scheme \cite{BB84}.
Moreover, since the averaged fidelity is
the same as that of the single state whose group orbit generates all
possible input states, the only feature that can depend on the
particular group in the following analysis is the irreducibility of
the representation $U_\phi^{\otimes M}\otimes (U_\phi^{\otimes N})^*$.
This is the same for the full rotation group ${\mathbf U(1)}$ and
for its subgroup ${\mathbf Z}_4$ for $N=1$ and $M\le 2$, whereas one
may expect a slight improvement of fidelities for larger $N$ and
$M$. 

\par Finally, we want to stress that the
method used in the present paper could be easily generalized to any
quality criteria---also called {\em cost-function}---different from
the averaged fidelity in Eq. (\ref{fidglob1}) and the single particle 
average fidelity (\ref{F}). As a matter of fact,
the averaging of the cost function will always lead to a an optimal
cloning that is covariant, as long as the cost-function is linear in
the cloning CP-map. However, it will not be necessarily
true that the optimal cloning map will have output state in the
symmetric tensor-product Hilbert space. 
Actually, also in the case of the single particle fidelity we found the
optimal maps starting from the assumption that the output state is supported
on the symmetric subspace. 

\section*{Acknowledgments}
This work has been jointly funded by the EC under the programs
EQUIP (Contract No. IST-1999-11053)  and
ATESIT (Contract No. IST-2000-29681), and by the Istituto Nazionale di
Fisica della Materia under the program PRA CLON.

\end{document}